\documentstyle[12pt]{article}
\newcommand{\text}{\rm}

\begin{document}

\title{{\bf \ Comments on Topologically Massive Gravity with Propagating Torsion}}
\author{{\bf J. L. Boldo$^{*}$, L. M. de Moraes}$^{*}${\bf \ and J. A.
Helay\"{e}l-Neto$^{*\dagger }$, }\vspace{2mm} \\
{\bf $^{\ast}$}CBPF, Centro Brasileiro{\bf \ }de Pesquisas F\'{\i}sicas \\
Rua Xavier Sigaud 150, 22290-180, Urca \\
Rio de Janeiro, Brazil\vspace{2mm}\\
{\bf $^{\dagger}$}UCP, Universidade Cat\'{o}lica de Petr\'{o}polis\\
Rua Bar\~{a}o do Amazonas 124, 25685-070\\
Petr\'{o}polis, Brazil\vspace{2mm}}
\maketitle

\begin{abstract}
We study and discuss some of the consequences of the inclusion of torsion in
3D Einstein-Chern-Simons gravity. Torsion may trigger the excitation of
non-physical modes in the spectrum. Higher-derivative terms are then added
up and tree-level unitarity is contemplated.

\setcounter{page}{0}\thispagestyle{empty}
\end{abstract}

\vfill\newpage\ \makeatother
\renewcommand{\theequation}{\thesection.\arabic{equation}} %
\renewcommand{\baselinestretch}{2}

\section{\ Introduction\-}

The concept of torsion as the antisymmetric part of the affine conection was
introduced by Cartan, in $1922.$ He recognized that torsion should be the
geometrical object related to the intrinsic angular momentum of matter, and
later, with the concept of spin, it was suggested that torsion should
mediate a contact interaction between spinning particles without however
propagating in matter-free space \cite{Kerlick, Hehl}. After that, a number
of authors reassessed the relation between torsion and spin in an attempt to
formulate General Relativity as a gauge theory, with the motivation that, at
a microscopic level, particles are classified by their mass and spin,
according to the Poincar\'{e} group \cite{Novello, De Sabbata}. In fact,
torsion can be minimally coupled to matter fields: the requirement that the
Dirac equation in a gravitational field preserve local invariance under
Lorentz transformations yields a direct interaction between torsion and
fermions. The observational consequences of a propagating torsion and its
interactions with matter have been reported in \cite{Caroll, Hammond,
Shapiro}.

Another important reason to consider torsion in gravity theory concerns the
discussion of unitarity and renormalizability for the linearized version of
the theory. It is well-known that, although standard Einstein gravity in
four-dimensional space-time seems satisfactory as a classical theory, its
quantum formulation presents a serious drawback: the non-renormalizability
problem. In order to circumvent this undesirable aspect, several
modifications in the theory have been proposed, such as the insertion of
higher-derivative terms, which generally leads to a non-unitary theory. In
this context, there appears string theory as the main promising framework to
overcome the problems of quantum gravity. As a matter of fact, torsion is
naturally incorporated in the string formulation of gravity. It is generally
suggested that the rank-3 antisymmetric field of the low-energy effective
Lagrangian dictated by string theory might be associated with torsion.
Furthermore, the spin-2 component of the torsion field may affect the
graviton propagator and can therefore modify the convergence properties of
the theory.

On the other hand, the interest in discussing gravity in three-dimensional
space-time has been receiving a great deal of attention \cite{Witten}. In
spite of the fact that planar Einstein theory has no dynamical degrees of
freedom, the introduction of a 3D topologically massive term \cite{Jackiw,
Gentil} yields a dynamical graviton and leads to a renormalizable model \cite
{Deser, Kleppe, Buch}. We believe that the introduction of new degrees of
freedom into this theory (here, those coming from the torsion field) could
open some trend for a better understanding of 3D quantum gravity in its full
potentiality: the metric and torsion quantum fluctuations mix up and lead to
a rich mass spectrum. Also, the power-counting renormalizability may be
improved once torsion is allowed, although unitarity demands some extra care.

In this paper, we introduce torsion in the framework of topologically
massive gravity theory via the usual ''minimal '' substitution, namely, by
replacing the Christoffel symbol, $\left\{ _{\mu \nu }^{\,\,\lambda
}\right\} $, by the Cartan affine connection. In the same way as in 4D
gravity, the curvature scalar in 3D does not provide any kinetic term for
the torsion components, so that their field equations are purely algebraic.
However, we see that dynamical contributions for the torsion arise from the
Chern-Simons term. The main purpose of our work is to employ the
spin-projection operator formalism to find out the propagators for the
metric and torsion fields. In view of the results we attain, a ghost-free 3D
gravity theory can be formulated once some constraints are imposed on the
parameters of the Lagrangians we discuss \cite{Nieuw1, Sezgin}.

Our work is organized as follows: in Section 2, we provide a short review on
Riemann-Cartan space-time. 3D torsion is decomposed into its
SO(1,2)-irreducible components. Then, one presents the full Lagrangian for
the theory in its linearized approximation. Section 3 is devoted to the
attainment of the propagators for some specific classes of Lagrangians and
to the check of the tree-level unitarity of the theory. Our Concluding
Comments are summarized in Section 4.

\section{Preliminaries}

We propose to carry out our analysis of the propagating torsion by starting
off from the following action for topologically massive gravity in three
dimensions: 
\begin{equation}
S=\int d^3x\left[ \sqrt{g}(a_1{\cal R}+a_2{\cal R}^2+a_3{\cal R}_{\mu \nu }%
{\cal R}^{\mu \nu })+a_4{\cal L}_{cs}\right] ,  \label{1}
\end{equation}
where 
\begin{equation}
{\cal L}_{cs}=\varepsilon ^{\mu \nu \kappa }\Gamma _{\kappa \lambda
}^{\,\,\,\,\,\,\,\,\rho }(\partial _\mu \Gamma _{\rho \nu
}^{\,\,\,\,\,\,\lambda }+\frac 23\Gamma _{\mu \sigma
}^{\,\,\,\,\,\,\,\lambda }\Gamma _{\nu \rho }^{\,\,\,\,\,\,\sigma })
\label{2}
\end{equation}
is the topological Chern-Simons term. $a_1$, $a_2$ and $a_3$ are free
coefficients, whereas $a_4$ is the parameter associated to Chern-Simons. One
should notice that the usual action for the cosmological constant and the
so-called translational Chern-Simons term \cite{Mielke} have not been
included in (\ref{1}).

Note that in three dimensions terms proportional to ${\cal R}_{\mu \nu
\kappa \lambda }{\cal R}^{\mu \nu \kappa \lambda }$ are not independent, due
to the fact that the curvature tensor can be written in terms of the Ricci
tensor (they have the same number of independent components).

We adopt here the Minkowski metric $\eta _{\mu \nu }=(+;-,-)$ and the Ricci
tensor ${\cal R}_{\mu \nu }={\cal R}_{\lambda \mu \nu
}^{\,\,\,\,\,\,\,\,\,\,\,\,\lambda }$.

Let us consider Riemann-Cartan space-time, where the affine connection is
non-symmetric in the first two indices and is no longer completely
determined by the metric \cite{Kerlick}. The antisymmetric part of the
affine connection is the torsion tensor: 
\begin{equation}
T_{\mu \nu }^{\,\,\,\,\,\,\,\lambda }=2\Gamma _{[\mu \nu
]}^{\,\,\,\,\,\,\,\,\,\,\lambda }=\Gamma _{\mu \nu }^{\,\,\,\,\,\,\,\lambda
}-\Gamma _{\nu \mu }^{\,\,\,\,\,\,\,\lambda }.  \label{3}
\end{equation}

In this space-time, the coefficients of the affine connection can be
expressed in terms of the metric and torsion, 
\begin{equation}
\Gamma _{\mu \nu }^{\,\,\,\,\,\,\,\lambda }=\left\{ _{\mu \nu }^{\,\,\lambda
}\right\} +K_{\mu \nu }^{\,\,\,\,\,\,\,\lambda },  \label{4}
\end{equation}
where $\left\{ _{\mu \nu }^{\,\,\lambda }\right\} $ is the Christoffel
symbol, which is completely determined by the metric: 
\begin{equation}
\left\{ _{\mu \nu }^{\,\,\lambda }\right\} =\frac 12g^{\kappa \lambda
}\left( \partial _\mu g_{\kappa \nu }+\partial _\nu g_{\mu \kappa }-\partial
_\kappa g_{\mu \nu }\right) ,  \label{5}
\end{equation}
and 
\begin{equation}
K_{\mu \nu }^{\,\,\,\,\,\,\,\lambda }=\frac 12\left( T_{\mu \nu
}^{\,\,\,\,\,\,\lambda }+T_{\,\,\,\mu \nu }^\lambda -T_{\nu \,\,\,\mu
}^{\,\,\lambda }\right)  \label{6}
\end{equation}
is the contortion tensor, which is antisymmetric in the last two indices.

In 3D, the torsion with its 9 degrees of freedom can be covariantly split
into its SO(1,2)-irreducible components: a trace part, $t_\mu \equiv T_{\mu
\nu }^{\,\,\,\,\,\,\nu }$, a totally antisymmetric part, $\varphi \equiv
\varepsilon ^{\mu \nu \lambda }T_{\mu \nu \lambda }$, and a traceless rank-2
symmetric tensor, $X_{\mu \nu }$. The splitting in the above components is
realized according to the following relation: 
\begin{equation}
T_{\mu \nu \lambda }=\varphi \varepsilon _{\mu \nu \lambda }+\frac 12(\eta
_{\nu \lambda }t_\mu -\eta _{\mu \lambda }t_\nu )+\varepsilon _{\mu \nu
\kappa }X_{\,\,\,\,\,\,\lambda }^\kappa .  \label{7}
\end{equation}
In order to read off the propagators and, consequently, the particle
spectrum of the theory, we linearize the metric-dependent part of the
Lagrangian by adopting the weak gravitational field approximation: 
\begin{equation}
g_{\mu \nu }(x)=\eta _{\mu \nu }+\kappa h_{\mu \nu }(x),  \label{8}
\end{equation}
where $\kappa $ stands for the gravitational coupling constant in 3D. $%
\kappa h_{\mu \nu }$ represents a small pertubation around flat Minkowski
space-time.

The action is invariant under general coordinate transformations, 
\begin{equation}
\delta h_{\mu \nu }(x)=\partial _\mu \xi _\nu (x)+\partial _\nu \xi _\mu (x),
\label{9}
\end{equation}
where $\xi _\mu $ is the gauge parameter. Therefore, it is necessary to fix
this gauge invariance in order to make the wave operator of the Lagrangian
non-singular. This is achieved by adding the De Donder gauge-fixing term, 
\begin{equation}
{\cal L}_{gf}=\frac 1{2\alpha }F_\mu F^\mu ,  \label{10}
\end{equation}
where 
\begin{equation}
F_\mu =\partial _\nu \left( h_{\,\,\,\,\mu }^\nu -\frac 12\delta
_{\,\,\,\,\mu }^\nu h\right) ,  \label{11}
\end{equation}
with $h\equiv h_\mu ^{\,\,\,\mu }$. No other gauge-fixing term is required,
since the torsion field behaves like a tensor under general coordinate
transformations.

First, let us confine ourselves to the study of the topologically massive
gravity without the higher derivative terms ${\cal R}^2$ and ${\cal R}_{\mu
\nu }{\cal R}^{\mu \nu }$ $\left( a_2=a_3=0\right) $. In this case, the
action is the sum of Einstein, Chern-Simons and gauge-fixing terms. So, by
decomposing the torsion into its irreducible components (\ref{7}), and
making use of the weak field approximation for the metric, the bilinear
terms can be collected as below: 
\begin{eqnarray}
\left. {\cal L}\right. &=&\frac{a_1\kappa ^2}2\left[ \frac 12h^{\mu \nu
}\Box h_{\mu \nu }-\frac 12h\Box h+h\partial _\mu \partial _\nu h^{\mu \nu
}-h^{\mu \nu }\partial _\mu \partial _\lambda \left. h^\lambda \right. _\nu
+\right.  \label{12} \\
&&+\left. \frac 1{\kappa ^2}(-3\varphi ^2-t_\mu t^\mu +2X_{\mu \nu }X^{\mu
\nu })\right] +  \nonumber \\
&&+\frac 1{2\alpha }\left[ -h^{\mu \nu }\partial _\mu \partial _\lambda
h_{\,\,\,\,\nu }^\lambda +h^{\mu \nu }\partial _\mu \partial _\nu h-\frac
14h\Box h\right] +  \nonumber \\
&&+a_4\left[ \frac{\kappa ^2}2\varepsilon ^{\mu \nu \lambda }(h_\lambda
^{\,\,\,\kappa }\Box \partial _\mu h_{\kappa \nu }-h_\lambda ^{\,\,\,\kappa
}\partial _\mu \partial _\kappa \partial _\rho h_{\,\,\,\,\nu }^\rho
)\right. +  \nonumber \\
&&-\frac{3\kappa }2\left( X^{\mu \nu }\Box h_{\mu \nu }+X^{\mu \nu }\partial
_\mu \partial _\nu h-2X^{\mu \nu }\partial _\mu \partial _\lambda
h_{\,\,\,\,\nu }^\lambda \right) +  \nonumber \\
&&-\left. \frac 12\varphi \partial _\mu t^\mu +\frac 14\varepsilon ^{\mu \nu
\lambda }t_\nu \partial _\mu t_\lambda +X^{\mu \nu }\partial _\mu t_\nu
-\varepsilon ^{\mu \nu \lambda }X_\lambda ^{\,\,\,\kappa }\partial _\mu
X_{\kappa \nu }\right] ,  \nonumber
\end{eqnarray}
where $a_1$ has dimension of mass, while the coefficient of the topological
term, $a_4,$ is dimensionless. Later on, suitable choice of these parameters
will be adopted in order to avoid tachyon and ghost modes in the spectrum.

\section{The Propagators}

We now rewrite the linearized Lagrangian (\ref{12}) in a more convenient
form, namely 
\begin{equation}
\left. {\cal L}\right. =\frac 12\sum_{\alpha \beta }\phi ^\alpha {\cal O}%
_{\alpha \beta }\phi ^\beta ,  \label{13}
\end{equation}
where $\phi ^\alpha =\left( h^{\mu \nu },X^{\mu \nu },t^\mu ,\varphi \right) 
$ and ${\cal O}_{\alpha \beta }$ is the wave operator. The propagators are
given by 
\begin{equation}
\left\langle 0\right| T\left[ \phi _\alpha \left( x\right) \phi _\beta
\left( y\right) \right] \left| 0\right\rangle =i\left( {\cal O}_{\alpha
\beta }\right) ^{-1}\delta ^3\left( x-y\right) .  \label{14}
\end{equation}

In order to invert the wave operator, we shall use an extension of the
spin-projection operator formalism introduced in \cite{Rivers, Nitsch},
where one needs to add now other four new operators coming from the
Chern-Simons and torsion terms. The operators for rank-2 symmetric tensors
in 3D are given by: 
\begin{eqnarray}
P_{\mu \nu ,\kappa \lambda }^{(2)} &=&\frac 12\left( \theta _{\mu \kappa
}\theta _{\nu \lambda }+\theta _{\mu \lambda }\theta _{\nu \kappa }\right)
-\frac 12\theta _{\mu \nu }\theta _{\kappa \lambda },  \label{15} \\
P_{\mu \nu ,\kappa \lambda }^{(1)} &=&\frac 12\left( \theta _{\mu \kappa
}\omega _{\nu \lambda }+\theta _{\mu \lambda }\omega _{\nu \kappa }+\theta
_{\nu \kappa }\omega _{\mu \lambda }+\theta _{\nu \lambda }\omega _{\mu
\kappa }\right) ,  \nonumber \\
P_{\mu \nu ,\kappa \lambda }^{(0-s)} &=&\frac 12\theta _{\mu \nu }\theta
_{\kappa \lambda },  \nonumber \\
P_{\mu \nu ,\kappa \lambda }^{(0-w)} &=&\omega _{\mu \nu }\omega _{\kappa
\lambda },  \nonumber \\
P_{\mu \nu ,\kappa \lambda }^{(0-sw)} &=&\frac 1{\sqrt{2}}\theta _{\mu \nu
}\omega _{\kappa \lambda },  \nonumber \\
P_{\mu \nu ,\kappa \lambda }^{(0-ws)} &=&\frac 1{\sqrt{2}}\theta _{\kappa
\lambda }\omega _{\mu \nu },  \nonumber
\end{eqnarray}
where $\theta _{\mu \nu }$ and $\omega _{\mu \nu }$ are respectively the
transverse and longitudinal projector operators for vectors. The other four
operators coming from the bilinear terms are 
\begin{eqnarray}
S_{\mu \nu ,\kappa \lambda } &=&\left( \varepsilon _{\alpha \kappa \mu
}\theta _{\nu \lambda }+\varepsilon _{\alpha \lambda \mu }\theta _{\nu
\kappa }+\varepsilon _{\alpha \kappa \nu }\theta _{\mu \lambda }+\varepsilon
_{\alpha \lambda \nu }\theta _{\mu \kappa }\right) \partial ^\alpha ,
\label{17} \\
R_{\mu \nu ,\kappa \lambda } &=&\left( \varepsilon _{\alpha \kappa \mu
}\omega _{\nu \lambda }+\varepsilon _{\alpha \lambda \mu }\omega _{\nu
\kappa }+\varepsilon _{\alpha \kappa \nu }\omega _{\mu \lambda }+\varepsilon
_{\alpha \lambda \nu }\omega _{\mu \kappa }\right) \partial ^\alpha , 
\nonumber \\
A_{\mu \nu } &=&\varepsilon _{\mu \nu \kappa }\partial ^\kappa ,  \nonumber
\\
E_{\mu \nu ,\kappa } &=&\eta _{\mu \kappa }\partial _\nu +\eta _{\nu \kappa
}\partial _\mu ;  \nonumber
\end{eqnarray}
$A,\,E$ and $R$ appear exclusively due to the inclusion of torsion.

We now present the relations between the above spin-projector operators. The
products between the usual Barnes-Rivers operators in D=3 are given by 
\begin{eqnarray}
P^{i-a}P^{j-b} &=&\delta ^{ij}\delta ^{ab}P^{j-b},  \label{17a} \\
P^{i-ab}P^{j-cd} &=&\delta ^{ij}\delta ^{bc}P^{j-a},  \nonumber \\
P^{i-a}P^{j-bc} &=&\delta ^{ij}\delta ^{ab}P^{j-ac},  \nonumber \\
P^{i-ab}P^{j-c} &=&\delta ^{ij}\delta ^{bc}P^{j-ac},  \nonumber
\end{eqnarray}
and satisfy the following tensorial identity: 
\begin{equation}
\left( P^{(2)}+P^{(1)}+P^{(0-s)}+P^{(0-w)}\right) _{\mu \nu ,\kappa \lambda
}=\frac 12\left( \eta _{\mu \kappa }\eta _{\nu \lambda }+\eta _{\mu \lambda
}\eta _{\nu \kappa }\right) .  \label{17b}
\end{equation}
We list below some of the non-trivial relations involving the new projectors
(\ref{17}): 
\begin{eqnarray}
SS &=&-16\Box P^{(2)},  \label{17c} \\
RR &=&-4\Box P^{(1)},  \nonumber \\
SP^{(2)} &=&S,  \nonumber \\
RP^{(1)} &=&R.  \nonumber
\end{eqnarray}

Thus, the wave operator can be decomposed in four sectors, namely

\begin{equation}
{\cal O}=\left( 
\begin{array}{cc}
M & N \\ 
P & Q
\end{array}
\right) ,  \label{18}
\end{equation}
where the blocks $M,N,P$ and $Q$ are all $2\times 2$-matrices. Their
non-trivial elements, expanded in terms of the spin projection operators,
are listed below: 
\begin{eqnarray}
(M_{11})_{\mu \nu ,\kappa \lambda } &=&\frac{a_1\kappa ^2\Box }2P_{\mu \nu
,\kappa \lambda }^{(2)}-\frac{\Box }{2\alpha }P_{\mu \nu ,\kappa \lambda
}^{(1)}-\frac{(a_1\kappa ^2\alpha +1)\Box }{2\alpha }P_{\mu \nu ,\kappa
\lambda }^{(0-s)}+  \label{19} \\
&&+\frac{\sqrt{2}\Box }{4\alpha }(P_{\mu \nu ,\kappa \lambda }^{(0-s\omega
)}+P_{\mu \nu ,\kappa \lambda }^{(0-\omega s)})-\frac{\Box }{4\alpha }P_{\mu
\nu ,\kappa \lambda }^{(0-\omega )}+\frac{a_4\kappa ^2\Box }4S_{\mu \nu
,\kappa \lambda },  \nonumber \\
(M_{12})_{\mu \nu ,\kappa \lambda } &=&(M_{21})_{\mu \nu ,\kappa \lambda }=-%
\frac{3a_4\kappa \Box }2(P_{\mu \nu ,\kappa \lambda }^{(2)}-P_{\mu \nu
,\kappa \lambda }^{(0-s)}),  \nonumber \\
(M_{22})_{\mu \nu ,\kappa \lambda } &=&2a_1\left( P_{\mu \nu ,\kappa \lambda
}^{(2)}+P_{\mu \nu ,\kappa \lambda }^{(1)}+P_{\mu \nu ,\kappa \lambda
}^{(0-s)}+P_{\mu \nu ,\kappa \lambda }^{(0-\omega )}\right) -\frac{a_4}%
2\left( S_{\mu \nu ,\kappa \lambda }+R_{\mu \nu ,\kappa \lambda }\right) , 
\nonumber
\end{eqnarray}

\begin{eqnarray}
(N_{21})_{\mu \nu ,\kappa } &=&\frac{a_4}2E_{\mu \nu ,\kappa },  \label{20}
\\
(P_{12})_{\mu ,\nu \kappa } &=&-\frac{a_4}2E_{\mu ,\nu \kappa },  \nonumber
\\
(Q_{11})_{\mu \nu } &=&-a_1(\theta _{\mu \nu }+\omega _{\mu \nu })+\frac{a_4}%
2A_{\mu \nu },  \nonumber \\
(Q_{12})_\mu &=&-(Q_{21})_\mu =\frac{a_4}2\partial _\mu ,  \nonumber \\
(Q_{22}) &=&-3a_1.  \nonumber
\end{eqnarray}

Before computing the inverse of the operator ${\cal O}$, it is perhaps
worthwhile to discuss an interesting specific case. Note that, as already
remarked, the massless Einstein theory in three-dimensional space-time is
not dynamical; however, the introduction of the topological mass term leads
to a dynamical theory with a physical propagating massive spin-two particle
(if we take $a_1>0$) \cite{Gentil}. In this context, we expect that the
Chern-Simons term provides dynamics for the torsion components too. Then,
let us now restrict the analysis to the Lagrangian of torsion in a flat
space-time background, in order to get a first idea on the behaviour of
torsion components from the point of view of the topological theory.
Eliminating the graviton terms (due to our choice of a flat space-time)
yields the Lagrangian density:

\begin{eqnarray}
\left. {\cal L}\right. &=&\frac{a_1}2(-3\varphi ^2-t_\mu t^\mu +2X_{\mu \nu
}X^{\mu \nu })+  \label{21} \\
&&+\frac{a_4}4(-2\varphi \partial _\mu t^\mu +\varepsilon ^{\mu \nu \lambda
}t_\nu \partial _\mu t_\lambda +  \nonumber \\
&&+4X^{\mu \nu }\partial _\mu t_\nu -4\varepsilon ^{\mu \nu \lambda
}X_\lambda ^{\,\,\,\kappa }\partial _\mu X_{\kappa \nu }).  \nonumber
\end{eqnarray}
Here, one notices that the curvature scalar, ${\cal R}$, provides only mass
terms for the fields, while the kinetic terms are all coming from the
Chern-Simons sector.

As there are no derivatives of the scalar field, $\varphi $, in the
Lagrangian, its equation of motion leads to 
\begin{equation}
\varphi =-\frac{a_4}{6a_1}\partial _\mu t^\mu .  \label{22}
\end{equation}
Inserting (\ref{22}) back into (\ref{21}), one obtains a Lagrangian written
exclusively in terms of $t_\mu $ and $X_{\mu \nu }$. Using the
spin-projector algebra, the propagators are readily obtained. Their explicit
form in momentum space are as follows:

\begin{eqnarray}
\left\langle XX\right\rangle &=&i\left\{ -\frac{a_1}{2\left(
a_4^2p^2-a_1^2\right) }P^{(2)}+\frac 1{2a_1}\left( P^{(1)}+P^{(0-s)}\right)
\right. +  \label{23} \\
&&-\left. \frac{\left( a_4^2p^2-12a_1^2\right) }{a_1\left(
5a_4^2p^2+12a_1^2\right) }P^{(0-\omega )}-\frac{a_4}{8\left(
a_4^2p^2-a_1^2\right) }S+\frac{a_4}{8a_1^2}R\right\} ,  \nonumber
\end{eqnarray}

\begin{equation}
\left\langle t_\mu t_\nu \right\rangle =\frac i{a_1}\theta _{\mu \nu }-\frac{%
ia_4}{2a_1^2}A_{\mu \nu }-\frac{12ia_1}{5a_4^2p^2+12a_1^2}\omega _{\mu \nu },
\label{24}
\end{equation}
where we have suppressed the indices in the field $X_{\mu \nu }$ and in the
projectors $P_{\mu \nu ,\kappa \lambda },$ $S_{\mu \nu ,\kappa \lambda }$
and $R_{\mu \nu ,\kappa \lambda }.$

From the above results, we see that this theory contains a massive
non-tachyonic pole in the spin-2 sector of $\left\langle X_{\mu \nu
}X_{\kappa \lambda }\right\rangle $. On the other hand, (\ref{24}) shows
that the $t_\mu $-field describes only the propagation of a longitudinal
tachyonic mode that does not couple to external conserved sources. The same
holds true for the mixed propagator $\left\langle X_{\mu \nu }t_\lambda
\right\rangle $, which we do not quote above.

The next step is to check the tree-level unitarity of the theory. This is
done by analysing the residues of the saturated propagator. The source with
which we are going to saturate the $\left\langle X_{\mu \nu }X_{\kappa
\lambda }\right\rangle $-propagator is compatible with the symmetries of the
theory and can be expanded in terms of a complete basis as follows: 
\begin{eqnarray}
\tau _{\mu \nu }(p) &=&a(p)p_\mu p_\nu +b(p)p_{(\mu }\stackrel{\sim }{p}%
_{\nu )}+c(p)p_{(\mu }\varepsilon _{\nu )}+  \label{25} \\
&&+d(p)\stackrel{\sim }{p}_\mu \stackrel{\sim }{p}_\nu +e(p)\stackrel{\sim }{%
p}_{(\mu }\varepsilon _{\nu )}+f(p)\varepsilon _{(\mu }\varepsilon _{\nu )},
\nonumber
\end{eqnarray}
where $p^\mu =(p^0,\stackrel{\rightarrow }{p})$, $\stackrel{\sim }{p}^\mu
=(p^0,-\stackrel{\rightarrow }{p})$, and $\varepsilon ^\mu =(0,\stackrel{%
\rightarrow }{\epsilon })$ are linearly independent vectors. They satisfy
the conditions: 
\begin{equation}
p^\mu \stackrel{\sim }{p}_\mu =(p^0)^2+(\stackrel{\rightarrow }{p})^2\neq 0,
\label{26}
\end{equation}
\begin{equation}
p^\mu \varepsilon _\mu =0,  \label{27}
\end{equation}
\begin{equation}
\varepsilon ^\mu \varepsilon _\mu =-1.  \label{28}
\end{equation}

The current-current transition amplitude in momentum space is the saturated
propagator 
\begin{equation}
{\cal A}=\tau ^{*\mu \nu }(p)\left\langle X_{\mu \nu }(p)X_{\kappa \lambda
}(p)\right\rangle \tau ^{\kappa \lambda }(p).  \label{29}
\end{equation}
Due to the source constraint $p_\mu \tau ^{\mu \nu }=0,$ only the transverse
projectors $P^{(2)}$, $P^{(0-s)}$ and $S$ can give non-vanishing
contributions to the amplitude. However, when we take the imaginary part of
the residue of the amplitude at the massive pole, $\mu ^2\equiv \left( \frac{%
a_1}{a_4}\right) ^2$, only the spin-2 sector of the propagator will
contribute. Then, we obtain the following result: 
\begin{equation}
I\,m({\cal R}es{\cal A})=\lim_{p^2\rightarrow \mu ^2}{\ -\frac
1{2a_1}(\left| \tau _{\mu \nu }\right| ^2-\frac 12\left| \tau _\mu
^{\,\,\,\mu }\right| ^2).}  \label{30}
\end{equation}
For a massive pole, the analysis can be done in the rest frame. In this case 
\begin{eqnarray}
p^\mu &=&\left( \mu ,0,0\right) ,  \label{31} \\
\stackrel{\sim }{p}^\mu &=&\left( \mu ,0,0\right) ,  \nonumber \\
\varepsilon ^\mu &=&(0,\stackrel{\rightarrow }{\epsilon }).  \nonumber
\end{eqnarray}
Thus, by expressing the sources in terms of this basis, (\ref{30}) becomes 
\begin{equation}
I\,m({\cal R}es{\cal A})=-\frac 1{4a_1}\left| f\right| ^2.  \label{32}
\end{equation}
Therefore, unlike the torsion-free theory, the sign of the Einstein part
must be negative, namely 
\[
a_1=-\frac 1{\kappa ^2}, 
\]
otherwise the massive mode would become ghost-like.

From the results (\ref{23}) and (\ref{32}), it follows that the Lagrangian (%
\ref{21}) describes the propagation of a ghost-free and non-tachyonic
torsion with spin-2 and mass $\mu ^2.$ In fact, as we shall see next, the
spin-2 component of torsion plays a central r\^{o}le here, in the sense that
only this component can affect the spin-2 sector of the graviton propagator.

We now turn back to invert the complete wave operator (\ref{18}). The
procedure is straightforward, but tedious. We have chosen to invert the
sectors $M$ and $Q,$ using the algebra of the projectors listed in (\ref{17a}%
-\ref{17c}). Afterwards, one must to invert $(M-NQ^{-1}P)$ and $%
(Q-PM^{-1}N). $ In so doing, the propagators are classified in four sectors,
namely

\begin{equation}
{\cal O}^{-1}=\left( 
\begin{tabular}{ll}
$X$ & $Y$ \\ 
$Z$ & $W$%
\end{tabular}
\right) ,  \label{33}
\end{equation}
where 
\begin{eqnarray}
X &=&(M-NQ^{-1}P)^{-1},  \label{34} \\
Z &=&-Q^{-1}PX,  \nonumber \\
W &=&(Q-PM^{-1}N)^{-1},  \nonumber \\
Y &=&-M^{-1}NW,  \nonumber
\end{eqnarray}
or explicitly, in terms of matrix elements $iX_{11}=\left\langle
hh\right\rangle ,iX_{12}=\left\langle hX\right\rangle ,...,$ $%
iW_{22}=\left\langle \varphi \varphi \right\rangle $. Now, let us present
the non-vanishing propagators in momentum space: 
\begin{eqnarray}
\left\langle hh\right\rangle  &=&i\left\{ -\frac{8a_1(5a_4^2p^2+4a_1^2)}{%
\kappa ^2p^2(a_4^2p^2-4a_1^2)^2}P^{(2)}+\frac{2\alpha }{p^2}P^{(1)}\right. +
\label{35} \\
&&-\frac{8a_1}{\kappa ^2p^2(9a_4^2p^2-4a_1^2)}P^{(0-s)}+  \nonumber \\
&&+\frac{4\left[ \kappa ^2\alpha \left( 9a_4^2p^2-4a_1^2\right) -4a_1\right] 
}{\kappa ^2p^2(9a_4^2p^2-4a_1^2)}P^{(0-w)}+  \nonumber \\
&&-\frac{8\sqrt{2}a_1}{\kappa ^2p^2(9a_4^2p^2-4a_1^2)}%
(P^{(0-sw)}+P^{(0-ws)})+  \nonumber \\
&&+\left. \frac{2a_4(a_4^2p^2+8a_1^2)}{\kappa ^2p^2(a_4^2p^2-4a_1^2)^2}%
S\right\} ,  \nonumber
\end{eqnarray}
\begin{eqnarray}
\left\langle Xh\right\rangle  &=&i\left\{ \frac{6a_4(a_4^2p^2+4a_1^2)}{%
\kappa (a_4^2p^2-4a_1^2)^2}P^{(2)}-\frac{6a_4}{\kappa (9a_4^2p^2-4a_1^2)}%
P^{(0-s)}\right. +  \label{36} \\
&&-\left. \frac{6\sqrt{2}a_4}{\kappa (9a_4^2p^2-4a_1^2)}P^{(0-sw)}-\frac{%
6a_4^2a_1}{\kappa (a_4^2p^2-4a_1^2)^2}S\right\} ,  \nonumber
\end{eqnarray}
\begin{eqnarray}
\left\langle XX\right\rangle  &=&i\left\{ -\frac{2a_1(7a_4^2p^2-4a_1^2)}{%
(a_4^2p^2-4a_1^2)^2}P^{(2)}+\frac 1{2a_1}P^{(1)}\right. +  \label{37} \\
&&-\frac{2a_1}{(9a_4^2p^2-4a_1^2)}P^{(0-s)}-\frac{(a_4^2p^2-12a_1^2)}{%
2a_1(5a_4^2p^2+12a_1^2)}P^{(0-w)}+  \nonumber \\
&&+\left. \frac{a_4(a_4^2p^2+2a_1^2)}{(a_4^2p^2-4a_1^2)^2}S+\frac{a_4}{8a_1^2%
}R\right\} ,  \nonumber
\end{eqnarray}
\begin{equation}
\left\langle X_{\mu \nu }t_\kappa \right\rangle =\frac{a_4}{4a_1^2}(\theta
_{\mu \kappa }p_\nu +\theta _{\nu \kappa }p_\mu )+\frac{6a_4}{%
5a_4^2p^2+12a_1^2}\omega _{\mu \nu }p_\kappa ,
\end{equation}
\begin{equation}
\left\langle \varphi X_{\mu \nu }\right\rangle =i\frac{%
a_4p^2(a_4^2p^2-2a_1^2)}{a_1^2(5a_4^2p^2+12a_1^2)}\omega _{\mu \nu },
\end{equation}
\begin{equation}
\left\langle t_\mu t_\nu \right\rangle =\frac i{a_1}\theta _{\mu \nu }-\frac{%
ia_4}{2a_1^2}A_{\mu \nu }-\frac{12ia_1}{5a_4^2p^2+12a_1^2}\omega _{\mu \nu },
\end{equation}
\begin{equation}
\left\langle t_\mu \varphi \right\rangle =\frac{2ia_4}{5a_4^2p^2+12a_1^2}%
p_\mu ,  \label{43}
\end{equation}
\begin{equation}
\left\langle \varphi \varphi \right\rangle =-\frac{2i(a_4^2p^2+2a_1^2)}{%
a_1\left( 5a_4^2p^2+12a_1^2\right) },  \label{44}
\end{equation}

Upon inspection of the poles of the propagators quoted above, we can state
the following results:

(i) there appears an undesirable double pole at $p^2=\left( \frac{2a_1}{a_4}%
\right) ^2$ located in the spin-2 sector of the $h_{\mu \nu }-X_{\kappa
\lambda }$ propagators (we shall handle this situation below);

(ii) there is a massive mode at $p^2=\left( \frac{2a_1}{3a_4}\right) ^2$ in
the spin-0 sector of the $h_{\mu \nu }-X_{\kappa \lambda }$ propagators;

(iii) a tachyonic pole shows up at $p^2=-\frac{12a_1^2}{5a_4^2}$ in the
longitudinal sector, $P^{(0-w)}$, of the $X_{\mu \nu }-t_\kappa $
propagators (it does not however contribute to the residue of the
current-current amplitude whenever the sources are transverse).

We now search for the constraints on the parameters of the theory that
follow from the requirement of having a positive-definite residue matrix at
the pole. Using (\ref{35}-\ref{37}), one obtains the following residue of
the saturated propagator at the pole $p^2=\left( \frac{2a_1}{3a_4}\right)
^2\equiv m_1^2:$

\begin{equation}
I\,m({\cal R}es{\cal A})=\lim_{p^2\rightarrow m_1^2}\left( 
\begin{array}{cc}
\tau ^{*} & \sigma ^{*}
\end{array}
\right) \left( 
\begin{array}{cc}
-\frac{8a_1}{p^2\kappa ^2} & -6a_4 \\ 
-6a_4 & -2a_1\kappa ^2
\end{array}
\right) P^{\left( 0-s\right) }\left( 
\begin{array}{c}
\tau \\ 
\sigma
\end{array}
\right) ,  \label{51}
\end{equation}
where we have redefined the $X_{\mu \nu }$-field$,X_{\mu \nu }^{^{\prime
}}=\kappa X_{\mu \nu },$ in order that $X_{\mu \nu }$ and $h_{\mu \nu }$
have the same dimension. $\tau $ is the external current associated to the
graviton field and $\sigma $ the one associated to the torsion. The above
residue matrix has one non-trivial positive eigenvalue (non-ghost) by
choosing $a_1<0$, which corresponds to a physical spin-0 mode.

Furthermore, we see that the inclusion of torsion into the
Einstein-Chern-Simons theory, by replacing the Christoffel symbol by the
Cartan connection, leads to a theory in which the spin-2 propagators contain
second order poles, and unitarity is consequently violated \cite{Nieuw2}. To
overcome this undesirable situation, we propose the introduction of the
following higher-derivative terms:

$a_2{\cal R}^2$ and $a_3{\cal R}^{\mu \nu }{\cal R}_{\mu \nu },$ as well as
the new coupling term, $a_5X^{\mu \nu }\widetilde{{\cal R}}_{\mu \nu },$
into the original Lagrangian, where the tilde in $\widetilde{{\cal R}}_{\mu
\nu }$ means that we are considering only the Riemannian part of this
tensor. The meaning of the last term will become clear later.

Confining ourselves, for simplicity, only to the $h_{\mu \nu }-X_{\kappa
\lambda }$ sector of the Lagrangian, since the other blocks do not affect
the spin-2 sector of the propagators, we can be concerned only with the
inversion of the M-sector of the wave operator, including of course the
contributions from the new coupling terms into its matrix elements. In this
case, the operator we shall deal with reads as follows: 
\begin{equation}
M=\left( 
\begin{array}{cc}
M_{11} & M_{12} \\ 
M_{21} & M_{22}
\end{array}
\right) ,  \label{46}
\end{equation}
where 
\begin{eqnarray}
M_{11} &=&M_{11,0}+\frac{a_3}2\kappa ^2\Box ^2P^{(2)}+\left( \frac
32a_3+4a_2\right) \kappa ^2\Box ^2P^{(0-s)},  \label{47} \\
M_{12} &=&M_{21}=M_{12,0}+\frac{a_3\kappa }4\Box S+\frac{a_5\kappa }4\Box
\left( P^{(2)}-P^{(0-s)}\right) ,  \nonumber \\
M_{22} &=&M_{22,0}-a_3\Box \left( 2P^{(2)}+P^{(1)}+P^{(0-s)}+\frac
12P^{(0-w)}\right) ,  \nonumber
\end{eqnarray}
where the $M_{..,0}$'s are the previous matrix elements coming from the
curvature escalar and Chern-Simons terms, listed in (\ref{19}). Proceeding
like in (\ref{34}), we may work out the $h_{\mu \nu }-X_{\kappa \lambda }$
propagators.

It now becomes clear the importance of the new coupling term $a_5X^{\mu \nu }%
\widetilde{{\cal R}}_{\mu \nu }$: it has the same operator structure as $%
M_{12,0}$; therefore, if we set $a_5=6a_4,$ and $a_3=0,$ the torsion field
decouples from the graviton and consequently they can be treated
independently. On the other hand, if $a_3\neq 0,$ $a_5=6a_4,$ and $a_1=0,$
one obtains the following results 
\begin{eqnarray}
\left\langle hh\right\rangle &=&i\left\{ \frac{2a_3}{p^2\kappa
^2(9a_3^2p^2-4a_4^2)}P^{(2)}+\frac{2\alpha }{p^2}P^{(1)}+\frac 2{p^4\kappa
^2(3a_3+8a_2)}P^{(0-s)}\right. +  \nonumber \\
&&+\frac{4\left[ \alpha \left( 3a_3+8a_2\right) \kappa ^2p^2+1\right] }{%
p^4\kappa ^2(3a_3+8a_2)}P^{(0-w)}+  \label{48} \\
&&+\left. \frac{2\sqrt{2}}{p^4\kappa ^2(3a_3+8a_2)}\left(
P^{(0-sw)+}P^{(0-ws)}\right) -\frac{(3a_3^2p^2-2a_4^2)}{2a_4p^4\kappa
^2(9a_3^2p^2-4a_4^2)}S\right\} ,  \nonumber
\end{eqnarray}
\begin{equation}
\left\langle Xh\right\rangle =-i\left\{ \frac{a_3^2}{a_4\kappa
(9a_3^2p^2-4a_4^2)}P^{(2)}+\frac{a_3}{2p^2\kappa (9a_3^2p^2-4a_4^2)}%
S\right\} ,  \label{49}
\end{equation}
\begin{eqnarray}
\left\langle XX\right\rangle &=&i\left\{ \frac{2a_3}{(9a_3^2p^2-4a_4^2)}%
P^{(2)}+\frac{a_3}{(a_3^2p^2-a_4^2)}P^{(1)}\right. +  \nonumber \\
&&+\frac 1{a_3p^2}P^{(0-s)}+\frac 2{a_3p^2}P^{(0-w)}+  \label{50} \\
&&+\left. \frac{a_4}{2p^2(a_3^2p^2-a_4^2)}R-\frac{(3a_3^2p^2-4a_4^2)}{%
a_4p^2(9a_3^2p^2-4a_4^2)}S\right\} .  \nonumber
\end{eqnarray}

Using (\ref{48}-\ref{50}), one obtains the following residue of the
saturated propagator at the pole $p^2=\left( \frac{2a_4}{3a_3}\right)
^2\equiv m_2^2:$

\begin{equation}
I\,m({\cal R}es{\cal A})=\lim_{p^2\rightarrow m_2^2}\left( 
\begin{array}{cc}
\tau ^{*} & \sigma ^{*}
\end{array}
\right) \left( 
\begin{array}{cc}
\frac{2a_3}{p^2\kappa ^2} & -\frac{a_3^2}{a_4} \\ 
-\frac{a_3^2}{a_4} & 2a_3\kappa ^2
\end{array}
\right) P^{\left( 2\right) }\left( 
\begin{array}{c}
\tau \\ 
\sigma
\end{array}
\right) ,  \label{51}
\end{equation}

From (\ref{51}), we obtain the following condition on the parameters of the
Lagrangian for not having a ghost at the massive pole: 
\begin{equation}
a_3=a_4\kappa ^2,\,\,\,\,a_4>0.  \label{52}
\end{equation}

Furthermore, we remark that the residue matrix of eq. (\ref{51}) provides
two positive eigenvalues, so that this theory propagates a massive spin-2
graviton along with a massive spin-2 quantum of the torsion.

As for the massless pole present in the $h_{\mu \nu }-X_{\kappa \lambda }$
propagators, it is harmless: indeed, like the case where torsion is absent,
it is non-dynamical, yielding a vanishing residue for the current-current
amplitude. So, the physical excitations propagated as gravitational degrees
of freedom are all massive.

\section{Concluding Comments}

The main purpose of our investigation was the assessment of the r\^{o}le
torsion may play in the framework of three-dimensional gravity in the
presence of the topological mass term.

Several peculiarities have been found out. For instance, the appearance of a
(gauge-independent) double pole in the spin-2 sector of the graviton and the
torsion propagators, which spoils the unitarity of the model. However, as we
have checked, the inclusion of higher powers of the curvature $\left( {\cal R%
}^2{\rm {\ and}}\text{{\rm \thinspace }}{\rm {\cal R}_{\mu \nu }{\cal R}%
^{\mu \nu }}\right) $ and a mixing term between the Ricci tensor and the
spin-2 part of the torsion may restore unitarity, for the parameters may be
so chosen that the double pole is suppressed.

We have contemplated the possibility of switching off the graviton degrees
of freedom and we have considered the propagation of torsion on a flat
background. In such a situation, no double pole shows up and we have checked
that the tree-level unitarity is ensured.

The three different spin sectors carried by the torsion tensor in 3D gravity
do not share common characteristics as long as the dynamics is concerned: in
the case in which only the Einstein-Hilbert and the Chern-Simons terms are
present, the scalar part, $\varphi ,$ behaves like an auxiliary field,
whereas the spin-1 sector, $t_\mu $, propagates only its longitudinal
component and the spin-2 sector, $X_{\mu \nu },$ is dynamical. We have not
here coupled matter to gravity with torsion. Neverthless, in so doing,
minimal coupling to fermions selects only the scalar component of the
torsion to couple directly to fermion bilinears \cite{Novo1}.

With the results we have got here, it might be worthwhile to extend our
analysis to 3D supergravity and try to understand how the dynamics of
torsion and its fermionic counterparts comes out.

Also, in view of the finiteness of pure gravity described by the
Chern-Simons term (\ref{2}), we should inquire, once we know how torsion
propagates and interacts, whether or not finiteness is kept whenever torsion
effects are included. This matter is now being considered and the results
shall be reported elsewhere \cite{Novo}.

\vspace{5mm}

{\Large {\bf Acknowledgements}}

We are grateful to Dr. I. L. Shapiro for helpful discussions and pertinent
comments. J. L. Boldo thanks CNPq and L. M. Moraes thanks CAPES for
financial support.

\end{document}